\documentclass[paper,notoc]{JHEP} 
\usepackage{epsfig}
\usepackage{amsmath}
\usepackage{amssymb}
\usepackage{cite}
\psfull

\def\cA{{\cal{A}}}
\def\cF{{\cal{F}}}
\def\cO#1{{\cal O}\left( {#1} \right)}
\def\asb{{\bar \alpha}_{\mbox{\scriptsize s}}}
\def\as{\alpha_{\mbox{\scriptsize s}}}

\def\eps{\epsilon}

\def\ca{C_A}

\def\qt{q_t}
\def\qti#1{q_{t,#1}}
\def\vqti#1{{\vec q}_{t,#1}}

\def\vq{{\vec q}}

\def\vk{{\vec k}}

\def\rhoti#1{\rho_{t,#1}}

\def\vrhoti#1{{\vec \rho}_{t,#1}}
\def\vrhoi#1{{\vec \rho}_{#1}}
\def\vrho{{\vec \rho}}

\title{Soft emissions and the equivalence of BFKL and CCFM final
  states} 

\author{G.P. Salam\thanks{This work was supported in
    part by the EU Fourth Framework Programme `Training and Mobility
    of Researchers', Network `Quantum Chromodynamics and the Deep
    Structure of Elementary Particles', contract FMRX-CT98-0194 (DG
    12-MIHT).}  \\  
  INFN, Sezione di
  Milano, Via Celoria 16, Milano, Italy\\
  E-mail: \email{G.Salam@mi.infn.it}}
  
\abstract{This article demonstrates that the BFKL and CCFM equations,
  despite their different physical content, lead to equivalent results
  for any final-state observable at leading single-logarithmic order.
  A novel and fundamental element is the treatment also of the soft
  ($z\to1$ divergent) part of the splitting function in the CCFM
  equation.}

\preprint{hep-ph/9902324\\
          Bicocca--FT--99--02\\
          January 1999}


\begin{document}

\section{Introduction}
\label{sec:intro}

For the study of the properties of small-$x$ deep inelastic scattering
collisions, it is desirable to sum leading logarithmic (LL) terms,
$(\as\ln x)^n$ of the perturbation series. This is generally done
using the BFKL equation \cite{BFKL}. Formally its validity is
guaranteed only for inclusive quantities, because its derivation
relies on the dominance of multi-Regge kinematics.

To guarantee leading logarithms of $x$ also for exclusive quantities,
it is necessary to take into account the QCD coherence and soft
radiation (that from the $z\to1$ divergent part of the splitting
function). This is done in the CCFM equation \cite{CCFMa,CCFMb}. For
inclusive quantities it can be demonstrated to give the same results
as the BFKL equation to LL level. For exclusive quantities this is not
necessarily the case.

A few years ago Marchesini \cite{Pino95} showed that multi-jet rates
do indeed differ in the BFKL and CCFM approaches: the latter involving
$(\as\ln^2x)^n$ factors, the former not. More recently Forshaw and
Sabio Vera \cite{FS} considered multi-jet rates with the extra
condition that the jets should be resolved (i.e.\ %
that their transverse momentum should be larger than a certain
resolution parameter $\mu_R$), and demonstrated to order $\as^3$ that
the BFKL and CCFM jet-rates were identical. This result was then
extended to all orders by Webber \cite{Webber}.

The results \cite{FS,Webber} were derived in the double-logarithmic
(DL) approximation, i.e.\ %
considering only powers of $\as$ that are accompanied by two
logarithms, at least one of which is a $\ln x$.  Additionally, in all
cases, the CCFM equation was used without the $z\to1$ divergent (soft)
part of the splitting function.

A number of questions arise. Firstly, taking the limit of $\mu_R\to0$
one should go from the result of Forshaw, Sabio Vera and Webber to
that of Marchesini --- the basic logarithmic structure being
different, it is not clear how this comes about. The resolution of
this puzzle comes through the consideration of formally subleading
terms $\as\ln^2 Q/\mu_R$, which must be resummed when taking the limit
$\mu_R\to0$, and which give a continuous transition from the case of
BFKL and CCFM results being equivalent, to that of their being
different. This is presented in section~\ref{sec:divDL}.

A second question is the importance of the $z\to1$ divergent part of
the splitting function in the CCFM equation. In previous studies this
has not been examined.  In section~\ref{sec:soft} it will be shown
that its inclusion leads to \textit{all} BFKL and CCFM final-state
properties being identical in the DL approximation.\footnote{This
  addresses also the issue, raised in \cite{Webber,FSW}, of whether
  differences might arise between BFKL and CCFM for
  \textit{correlations} between resolved jets.}  It is also
demonstrated that no potentially dangerous $\as\ln^2\mu_R$ terms arise
anymore (either in CCFM or BFKL), except some associated with the end
of the branching chain.

Finally in section~\ref{sec:SingLog} the above result is extended to
single logarithmic accuracy (i.e.\ %
terms where all powers of $\as$ are accompanied by a $\ln x$ factor,
but not necessarily by any other logarithm). For an actual calculation
of final-state properties to single-logarithmic accuracy, the reader
is referred to \cite{EW}.

\section{Dividing up the double-logarithmic physics}
\label{sec:divDL}

\subsection{BFKL}

\FIGURE{
    \input{kinb.pstex_t}
    \caption{kinematics.}
    \label{fig:kin}}

Let us start with the BFKL equation. We will consider ladders with
kinematics labelled as in figure~\ref{fig:kin}. The exchanged gluon
$i$ has longitudinal momentum fraction $x_i$, and transverse momentum
$k_i$. We define also $z_i=x_i/x_{i-1}$. The transverse momentum
coming into the chain is taken to be zero. Emitted gluons have
transverse momentum $\qti{i}$ and longitudinal momentum fraction
$x_{i-1}(1-z_i)$. 

The unintegrated gluon density is given by
\begin{multline}
\label{eq:BFKLgd}
  \cF(x,k) = \sum_{n=0}^\infty \prod_{i=1}^n \left(\asb
    \int\frac{dz_i}{z_i} \int_\mu\frac{d^2\vqti{i}}{\pi\qti{i}^2} 
  \Delta(z_i,k_i)\right) \\
\delta\!\!\left(x - \prod_{i=1}^n z_i \right)
\delta^2\!\!\left(\vk + \sum_{i=1}^n \vqti{i} \right),
\end{multline}
where $n$ is the number of emissions, $\mu$ is a collinear cutoff and
$\asb = \as \ca/\pi$. For $n=0$, the product of $z_i$'s should be
interpreted as being equal to $1$, so that the initial condition is
$\delta(x-1) \delta^2(\vk)$. The virtual corrections are contained in
the form factor $\Delta(z,k)$:
\begin{equation}
  \ln \Delta(z,k) =  -2\asb \ln \frac1z \, \ln \frac{k}{\mu}\,,
\end{equation}
and the $k_i$ are given by 
\begin{equation}
  \label{eq:kisum}
  \vk_i = -\sum_{j=1}^i \vqti{i}\,, \qquad k = k_n\,.
\end{equation}

Since we have two large quantities, $\ln k/\mu$ and $\ln 1/x$, it is of
interest to carry out a double logarithmic expansion, i.e.\ %
concentrating on terms where each power of $\as$ is accompanied by two
logarithms (in the BFKL case always the product $\ln 1/x \,\ln k/\mu$).

Each branching is guaranteed to give a factor $\as \ln1/x$. To obtain
an extra (transverse) logarithm, the branching has to be sensitive to
a ratio of transverse scales. The two such kinds of branching are
those that increase the exchanged momentum $k$ ($k$-changing
emissions), $k_i\simeq \qti{i} \gg k_{i-1}$, which give double
logarithms in the total cross section; and those that don't change it
at all ($k$-conserving emissions), $k_i\simeq k_{i-1} \gg \qti{i}$,
giving double logarithms only in the final state.

All remaining kinds of branching give just a factor of $\as \ln 1/x$,
and so can be neglected to double-logarithmic accuracy. Thus,
\eqref{eq:BFKLgd} becomes
\begin{multline}
\label{eq:BFKLdl}
  \cF(x,k) = \sum_{n=0}^\infty \prod_{i=1}^n \left(\asb
    \int\frac{dz_i}{z_i} \int_\mu\frac{d^2\vqti{i}}{\pi\qti{i}^2} 
  \Delta(z_i,k_i)\right) \\ \cdot
\delta\!\!\left(x - \prod_{i=1}^n z_i \right)
\delta^2\!\!\left(\vk + \max(\vqti{1},\ldots,\vqti{n})\right)\,,
\end{multline}
where the $k_i$ are now defined as
\begin{equation}
  \label{eq:kimax}
  k_i = -\max(\vqti{1},\ldots,\vqti{i})\,.
\end{equation}
We then note that 
\begin{equation}
\label{eq:BFKLone}
  1 = 
  \sum_{m=0}^\infty \,\prod_{\ell=1}^{m}
  \left(\asb \int \frac{d\zeta_\ell}{\zeta_\ell} \int_\mu^k
    \frac{d^2\vrhoti{\ell}}{\pi\rhoti{\ell}^2} 
    \Delta(\zeta_\ell,k)
  \right) \Delta(z/\zeta_\Pi, k)
  \,\Theta\!\!\left(\zeta_\Pi - z \right),
\end{equation}
where $z$ and $k$ can take any value and
\begin{subequations}
\label{eq:zetaPi}
\begin{align}
  m &= 0 :  \qquad  \zeta_\Pi = 1\,, \\
  m &\ge 1: \qquad \zeta_\Pi =\prod_{\ell=1}^m \zeta_\ell\,. 
\end{align}
\end{subequations}
The relation \eqref{eq:BFKLone} is easily verified. Its significance
is that $k$-conserving emissions are `probability conserving': they are
exactly compensated for by the virtual corrections.

Let us now rewrite \eqref{eq:BFKLdl} with $k$-changing emissions and
$k$-conserving emissions in separate sums,
\begin{multline}
\label{eq:BFKLfactored}
  \cF(x,k) = \sum_{n=0}^\infty 
\delta^2\!\!\left(\vk +  \vqti{n} \right)
\prod_{i=1}^n \Bigg[ \left(\asb
    \int\frac{dz_i}{z_i} \int_{k_{i-1}}
    \frac{d^2\vqti{i}}{\pi\qti{i}^2}  
  \right) 
\delta\!\!\left(x - \prod_{i=1}^n z_i \right)
\\ \cdot
  \sum_{m=0}^\infty \,\prod_{\ell=1}^{m}
  \left(\asb \int \frac{d\zeta_\ell}{\zeta_\ell} \int_\mu^{k_i}
    \frac{d^2\vrhoti{\ell}}{\pi\rhoti{\ell}^2} 
    \Delta(\zeta_\ell,k_i)
  \right) \Delta(z_i/\zeta_{\Pi_i}, k_i)
  \,\Theta\!\!\left(\zeta_{\Pi_i} - z_i \right) \Bigg],
\end{multline}
\noindent where $\zeta_{\Pi_i}$ is defined in analogy with \eqref{eq:zetaPi},
$k_0=\mu$ and $k_i=\qti{i}$.  Exploiting \eqref{eq:BFKLone} to
eliminate the second line, one sees that the double-logarithmic BFKL
cross section is given by
\begin{equation}
\label{eq:BFKLdglap}
  \cF(x,k) = \sum_{n=0}^\infty \prod_{i=1}^n \left(\asb
    \int\frac{dz_i}{z_i} \int_{k_{i-1}}
    \frac{d^2\vqti{i}}{\pi\qti{i}^2}  
  \right) 
\delta\!\!\left(x - \prod_{i=1}^n z_i \right)
\delta^2\!\!\left(\vk +  \vqti{n} \right),
\end{equation}
which is just the double-logarithmic DGLAP (ordered chain)
\cite{DGLAP} result for the gluon density. To obtain the full BFKL
final-state prediction one then `dresses' the ordered chain
(figure~\ref{fig:dressing}) by putting back in a set of $k$-conserving
emissions after each $k$-changing emission,
i.e.\ %
by replacing the second line of \eqref{eq:BFKLfactored}.
\FIGURE{    \input{dressing.pstex_t}
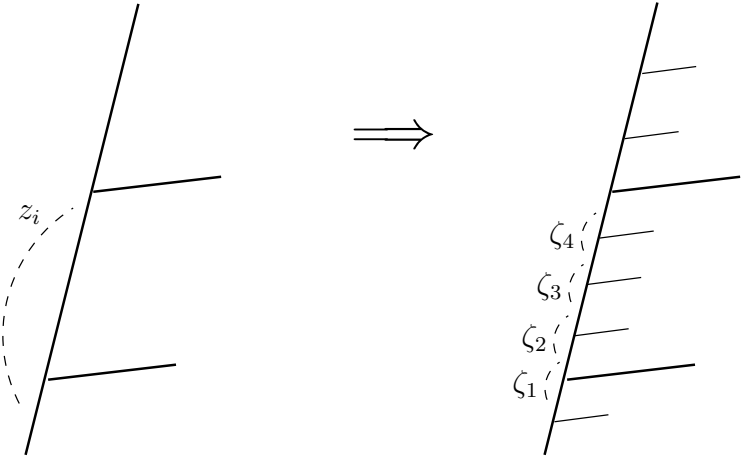
    \caption{Adding low-$\qt$ emissions to a strongly ordered chain:
      $\zeta_1\cdots\zeta_4 = z_i$.}
    \label{fig:dressing}}

The rapidity of each emission $i$ is given by
\begin{equation}
  \eta_i = \ln \frac{\qti{i}}{x_{i-1} (1-z_i)}\,.
\end{equation}
The procedure of removing (or adding back in) the $k$-conserving
emissions modifies the $z_i$'s for the $k$-changing emissions, but not
the $x_{i-1}$ values.  Since $z_i$ is anyway much less than $1$,
$1-z_i$ changes by a negligible (next-to-leading) amount --- so strong
ordering ensures that rapidities are essentially unaffected by the
removal and insertion of a subset of emissions, allowing one to safely 
use this technique for examining properties of the final state.

\subsection{CCFM}

The CCFM gluon density is given by
\begin{multline}
\label{eq:CCFMgd}
\cA(x,k,p) = \sum_{n=0}^\infty \prod_{i=1}^n \left(\asb
  \int\frac{dz_i}{z_i} \int \frac{d^2\vq_i}{\pi q_i^2}
  \,\Delta(z_i,k_i,q_i) \,\Theta(q_i - z_{i-1}
  q_{i-1})\,\Theta(\qti{i} - \mu)\right) \\ \cdot \Theta(p-z_n q_n)\,
\delta\!\left(x - \prod_{i=1}^n z_i \right) \delta^2\!\!\left(\vk +
  \sum_{i=1}^n \vqti{i} \right).
\end{multline}
Angular ordering is embodied by the factors $\Theta(q_i - z_{i-1}
q_{i-1})$ (absent for $i=1$). A maximum angle is introduced through
the dependence on the third variable, $p$. The $k_i$ are defined as in 
\eqref{eq:kisum}.

Relative to the BFKL equation, the virtual corrections differ so as to
take into account the angular ordering, and are given by the
non-Sudakov form factor $\Delta(z,k,q)$:
\begin{subequations}
\begin{align}
  \ln \Delta(z,k,q) &= 
 -2\asb \int_z^1 \frac{d\zeta}{\zeta} \int \frac{d\rho}{\rho}
  \,\Theta(k-\rho)  \, \Theta(\rho -\zeta q) \,\Theta(\rho - \mu)\\ &= 
 -\asb \left(\ln^2 \frac1z + 2\ln \frac1z\ln
    \frac{k}{q}\right), \qquad q < k,\;\; zq > \mu\,.
\end{align}
\end{subequations}
It is to be noted that $q$ is a rescaled transverse momentum,
$q=\qt/(1-z)$. Since we have only the $1/z$ part of the splitting
function, we work in the limit of $z \ll 1$ and so the difference
between $q$ and $\qt$ can be neglected.\footnote{As can be any factors
  of $1-z$. However when we introduce soft emissions, later on, it
  will be vital to retain their associated $1-z$ factors.}

As in the BFKL case, when considering the DL limit, we can replace
$\sum \vqti{i}$ in the $\delta$-function by the largest of the
$\qti{i}$:
\begin{multline}
\label{eq:CCFMdl}
  \cA(x,k,p) = \sum_{n=0}^\infty \prod_{i=1}^n \left(\asb
    \int\frac{dz_i}{z_i} \int\frac{d^2\vq_i}{\pi q_i^2} \,
  \Delta(z_i,k_i,q_i) \,\Theta(q_i - z_{i-1} q_{i-1})\,\Theta(\qti{i} -
  \mu)\right) \\ \cdot \Theta(p-z_n q_n)\,
 \delta\!\left(x - \prod_{i=1}^n z_i \right)
\delta^2\!\!\left(\vk + \max(\vqti{1},\ldots,\vqti{n})\right),
\end{multline}
with the $k_i$ now defined as in \eqref{eq:kimax}.
The next step is to note that 
\begin{multline}
\label{eq:CCFMone}
  1 = 
  \sum_{m=0}^\infty \,\prod_{\ell=1}^{m}
  \left(\asb \int \frac{d\zeta_\ell}{\zeta_\ell} \int^k
    \frac{d^2\vrhoi{\ell}}{\pi\rho_{\ell}^2} 
    \,\Delta(\zeta_\ell,k,\rho_\ell) \,\Theta(\rho_\ell -
    \zeta_{\ell-1} \rho_{\ell-1})  
    \,\Theta(\rho_{\ell} - \mu)
  \right) \\ \cdot \Delta(z/\zeta_\Pi, k, \rho_0)
  \,\Theta\!\!\left(\zeta_\Pi - z \right),
\end{multline}
where $\zeta_\Pi$ is defined as in the BFKL case \eqref{eq:zetaPi},
$\zeta_0 = z / \zeta_\Pi$, and $z$, $k$ and $\rho_0 \le k$ can take
any values; $\rho$ is a rescaled momentum, and $\rho_t = (1-\zeta)
\rho$ is the corresponding transverse momentum (at our accuracy, they
are not distinguishable for hard emissions, hence we are allowed to
write $\Theta(\rho_\ell-\mu)$ rather than the slightly more correct
$\Theta(\rhoti{\ell}-\mu)$). To demonstrate the relation
\eqref{eq:CCFMone} term by term is quite difficult.  Instead one can
see that it holds because it relates to a probabilistic branching ---
and the total probability of all possible states is $1$.

As in the BFKL case, the aim is now to show that the cross section is
determined only by the $k$-changing emissions. To be able to perform
the sum \eqref{eq:CCFMone} after each $k$-changing emission, there are
three conditions. Let us label a pair of successive $k$-changing
emissions as $a$ and $b$. The first condition is $\rho_0 \le k_a$, and
it is satisfied because $\rho_0\equiv q_a \simeq k_a$.  The second
condition is that the angular ordering of the next $k$-changing
emission should not cut out any piece of the sum \eqref{eq:CCFMone}:
this is guaranteed since $q_{b} > k_a$, so that the angular ordering
condition $z_a/\zeta_\Pi \rho_m < q_{b}$ is by definition satisfied
(recall that $\rho_m < k_a$). Finally, we want to carry out the
complete sum \eqref{eq:CCFMone} after the last $k$-changing emission
--- for this to be possible we require that $p\ge k$.

Therefore one can extract the $k$-conserving emissions from the
calculation of the cross section and obtain
\begin{multline}
\label{eq:CCFMdglap}
  \cA(x,k,p\ge k) = \sum_{n=0}^\infty \prod_{i=1}^n \left(\asb
    \int\frac{dz_i}{z_i} \int_{\max(k_{i-1},\mu)}
    \frac{d^2\vqti{i}}{\pi\qti{i}^2}  
  \right) \\
\Theta(p-z_n q_n) \, \delta\!\!\left(x - \prod_{i=1}^n z_i \right)
\delta^2\!\!\left(\vk +  \vqti{n} \right).
\end{multline}
This is the same result as in the BFKL case \eqref{eq:BFKLdglap}.

To obtain the correct final state one must reintroduce after each
$k$-changing emission a set of $k$-conserving emissions as given by
\eqref{eq:CCFMone}.

\subsection{BFKL and CCFM final states}
\label{sec:BCfs}

\FIGURE{    \input{phase-space.pstex_t}
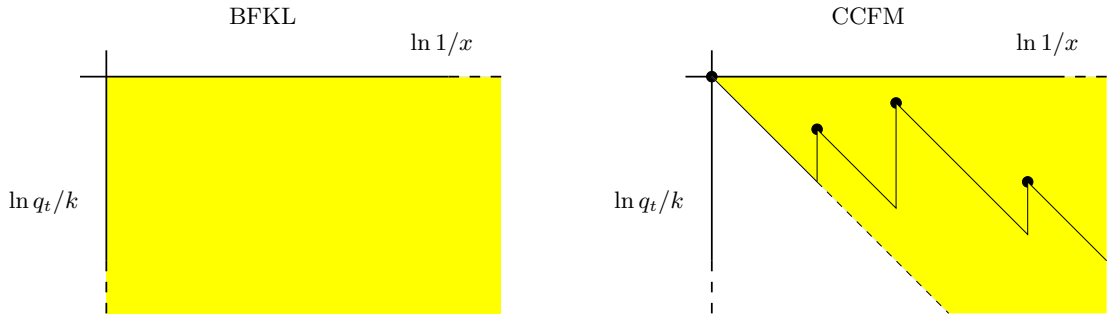
    \caption{Phase-space available in the BFKL and CCFM cases. The
      shaded-area is the phase-space available to the first emission.
      In the CCFM case the black circles represent further emissions,
      and the lines the corresponding subsequent delimitation of the
      phase-space for future emissions.}
    \label{fig:phasespace}}

Since the $k$-changing parts of the BFKL and CCFM equations are
identical, it suffices, at least for now, to consider the
$k$-conserving parts, \eqref{eq:BFKLone} and \eqref{eq:CCFMone}
respectively.

The BFKL case is extremely straightforward. The emission of a gluon is
a Poissonian type process, so that emissions are independent and
\begin{equation}
\label{eq:bfkldens}
\left \langle \frac{dn}{d\ln \qt \, d \ln 1/x} \right\rangle = 2\asb\,,
\end{equation}
where $x$ is the longitudinal momentum fraction of the emission.

The CCFM case is more complex. The phase-space for the first emission
differs from that in the BFKL case by a triangular region. The area of
the difference is proportional to $\ln^2 x$, and given that one
expects a number of emissions per unit area to be proportional to
$\as$, one sees immediately the well-known result \cite{Pino95} that
there is a difference between the BFKL and CCFM predictions at the
level of $\as \ln^2x$.

If one introduces a resolvability cutoff, as done for example by
Forshaw and Sabio Vera \cite{FS} (let $\mu_R$ be the resolution
scale), then the difference in available phase-space for a single
emission can never be larger than $\ln^2 k/\mu_R$: one loses the
double logarithm in $x$, and the difference between the BFKL and CCFM
results is now subleading, $\as\ln^2 k/\mu_R$.

This is the situation for a single CCFM emission. It is also possible
to study the asymptotic properties of the final state.

The number of emissions with (rescaled) transverse momentum $q_g$ and
momentum fraction $x_g$ contributing to the evolution of the gluon
density to a point $x,k$ is given by
\begin{multline}
\label{eq:CCFMdn}
 \frac{d n(x,k, x_g, q_g)}{d \ln x_g \, d^2 \vq_g} \cA(x,k,k) = 
 \sum_{n=0}^\infty \prod_{i=1}^n \bigg(\asb
    \int\frac{dz_i}{z_i} \int \frac{d^2\vq_i}{\pi q_i^2} 
  \,\Delta(z_i,k_i,q_i) \,\Theta(q_i - z_{i-1}
  q_{i-1}) \\ \cdot \,\Theta(\qti{i} - 
  \mu)\bigg) 
\frac{\asb}{\pi} \,\Theta\!\!\left(z_\Pi - x/x_g \right)
 \cA(x_g,k_g,q_g) \Delta(z_0, k_0, q_g).
\end{multline}
where $z_\Pi$ is defined in analogy with $\zeta_\Pi$ in
\eqref{eq:zetaPi} and
\begin{equation}
\vk_i = \vk + \sum_{j=i+1}^n \vqti{j}, \qquad \vk_g = \vk_0 +\vqti{g}, 
\qquad z_0 = \frac{x}{x_g z_\Pi}, \qquad q_0 \equiv q_g\,.
\end{equation}
To simplify the notation, $n$ has been taken for $p=k$. For $x\ll
x_g$, the dependence on $p$ would in any case be entirely contained in
the $\cA(x,k,p)$ factor.

In the case of nothing but $k$-conserving hard emissions
\eqref{eq:CCFMdn} simplifies to
\begin{multline}
\label{eq:CCFMdnkcons}
 \frac{d n(x,k, x_g, q_g)}{d \ln x_g \, d \ln q_g} \cA(x,k,k) = 
 2\asb \cA(x_g,k,q_g) \sum_{n=0}^\infty \prod_{i=1}^n \bigg(\asb
    \int\frac{dz_i}{z_i} \int^k \frac{d^2\vq_i}{\pi q_i^2} 
  \,\Delta(z_i,k,q_i)\\\cdot \,\Theta(q_i - z_{i-1}
  q_{i-1})\,\Theta(q_i - 
  \mu)\bigg)
 \,\Theta\!\!\left(1 - z_0 \right)
 \Delta(z_0, k, q_g).
\end{multline}
Making use of \eqref{eq:CCFMone}, and the fact that in the absence of
$k$-changing emissions $\cA(x,k,k)$ is independent of $x$, we ``evaluate''
all the integrals, and drop the explicit $x$ and $k$ dependence in $n$
to obtain
\begin{equation}
 \frac{d n(x_g, q_g/k)}{d \ln x_g \, d \ln q_g} = 
 2\asb \frac{\cA(x_g,k,q_g)}{\cA(x_g,k,k)} 
\end{equation}
It has been shown in \cite{CCFMa,BMSS97} that asymptotically,
\begin{equation}
   \frac{\cA(x,k,q)}{\cA(x,k,k)} \simeq e^{-\asb \ln^2 k/q},
   \qquad q \ll k\,.
\end{equation}
In the absence of $k$-changing emissions, this form is an exact
eigenfunction of the DL evolution equation.
Hence we arrive at the result that
\begin{equation}
 \label{eq:CCFMdnFinal}
 \frac{d n(x_g, \qti{g}/k)}{d \ln x_g \, d \ln \qti{g}} = 2\asb
 e^{-\asb \ln^2 k/\qti{g}}\,,
\end{equation}
\noindent where use has been made of the equivalence between $\qt$ and 
$q$ at this accuracy.  The first thing to note is that it differs from
the BFKL result by subleading terms $\as (\as \ln^2k/\qti{g})^n$. By
looking at the differential distribution with respect to $\ln
\qti{g}$, we have effectively introduced a resolvability cutoff, and
as discussed earlier in the single emission case, this ensures the
absence of double logarithms of $x$.

If on the other hand we integrate over all $\qti{g}$, we have that
\begin{equation}
 \frac{d n(x_g)}{d \ln x_g} = 2\asb \int^k \frac{d\qti{g}}{\qti{g}} 
 \frac{d n(x_g, \qti{g}/k)}{d \ln x_g \, d \ln  \qti{g}}
  = \sqrt{\pi\asb}\,.
\end{equation}
The total number of emissions in a 
region  $1> x_g > x$ is then
\begin{equation}
  n(x_g > x) \simeq \sqrt{\as \ln^2 x}
\end{equation}
which is a double-logarithm in $x$.

So there is a close connection between double logarithms in $x$, and
(formally subleading) double logarithms in $\qt$. The latter must be
retained if one wants to take the limit of $\mu_R\to0$. To put it in a
different way, if one is interested in values of $\qti{g}$
sufficiently low that that $\asb \ln^2 k/\qti{g} \gtrsim 1$, then
subleading double logarithms in $q$ must be resummed, and lead to a
significant difference between the CCFM ($z\to0$ divergent part only)
and BFKL predictions.

\section{CCFM with soft emissions}
\label{sec:soft}

\subsection{Factoring out soft emissions}

So far we have examined the CCFM equation with only the $1/z$ part of
the splitting function.  For brevity, emissions produced by the
$1/(1-z)$ part of the splitting function will be referred to as `soft'
emissions --- they being soft relative to the exchanged gluon off
which they are emitted.

The version of the CCFM equation with soft emissions has been examined
relatively little, apart from its use for phenomenology in the SMALLX
program \cite{SMALLX}, most recently studied in \cite{GBetal}. Part
of the reason is the considerable technical difficulty involved,
partly also uncertainty about exactly how the soft emissions are best
implemented. 

In the leading-logarithmic limit it turns out that these difficulties
disappear, or become irrelevant, since they are mostly related to
subleading issues.

The branching equation including the soft emissions is 
\begin{multline}
\label{eq:softgd}
  \cA(x,k,p) = \sum_{n=0}^\infty \prod_{i=1}^n \left(\asb
    \int dz_i     \frac{d^2\vq_i}{\pi q_i^2}  
    \bigg(\frac1{z_i} + \frac1{1-z_i}\right)
  \frac{\Delta_S(q_i)}{\Delta_S(z_{i-1} q_{i-1})} 
  \,\Delta(z_i,k_i,q_i) \\ \cdot
  \,\Theta(q_i - z_{i-1} q_{i-1})\,\Theta(\qti{i} -
  \mu)\bigg)
\delta\!\left(x - \prod_{i=1}^n z_i \right)
\delta^2\!\!\left(\vk + \sum_{i=1}^n \vqti{i} \right)
  \frac{\Delta_S(p)}{\Delta_S(z_{n} q_{n})} \Theta(p-z_n q_n) ,
\end{multline}
where $\Delta_S$, known as the Sudakov form factor, is 
given by 
\begin{equation}
  \label{eq:deltaS}
  \ln \Delta_S(p) = -2\asb \int 
  \frac{dz}{1-z} \int \frac{dq}{q}\,
    \Theta(\qt - \mu) \,\Theta(p-q) \,, \qquad p> \mu\,,
\end{equation}
and $z_0 q_0 = \mu$. The $k_i$ are defined as in \eqref{eq:kisum}.

In analogy with what was done above, one can take the double
logarithmic limit of \eqref{eq:softgd} to obtain
\begin{multline}
\label{eq:softdl}
  \cA(x,k,p) = \sum_{n=0}^\infty \prod_{i=1}^n \bigg(\asb
    \int dz_i     \frac{d^2\vq_i}{\pi q_i^2}  
    \left(\frac1{z_i} + \frac1{1-z_i}\right)
  \frac{\Delta_S(q_i) \,\Delta(z_i,k_i,q_i)}{\Delta_S(z_{i-1} q_{i-1})} 
  \,\Theta(q_i - z_{i-1} q_{i-1})   \\ \cdot
\,\Theta(\qti{i} -
  \mu)\bigg)
\delta\!\left(x - \prod_{i=1}^n z_i \right)
\delta^2\!\!\left(\vk + \max(\vqti{1},\ldots,\vqti{n})\right).
  \frac{\Delta_S(p)}{\Delta_S(z_{n} q_{n})} \Theta(p-z_n q_n)\,.
\end{multline}
The $k_i$ must be redefined as in \eqref{eq:kimax}.

We now examine how to separate the soft emissions from the
others, much in the same way as was done for separating the
$k$-conserving emissions from the $k$-changing emissions.

First, for all soft emissions we set $z_i=1$ everywhere except in the
$1/(1-z_i)$ factor (remembering that $\Delta(1,k,q)=1$). The error
that arises from this cannot be larger than next-to-leading, and will
not be enhanced by double logarithms.  We then observe that
\begin{multline}
\label{eq:softone}
  1 = \sum_{m=0}^\infty \prod_{\ell=1}^m \bigg(\asb
    \int    \frac{d^2\vrho_\ell}{\pi \rho_\ell^2}  
    \frac{d\zeta_\ell}{1-\zeta_\ell}
  \frac{\Delta_S(\rho_\ell)}{\Delta_S( \rho_{\ell-1})} 
  \,\Theta(\rho_\ell - \rho_{\ell-1})\,\Theta(\rhoti{\ell} -
  \mu)\bigg) \\ \cdot
  \frac{\Delta_S(P)}{\Delta_S(\rho_{m})} \Theta(P- \rho_m) ,
\end{multline}
for any value of $P > \rho_0$.  When inserting this sum between two
`hard' emissions (say $q_a$ and $q_b$), it should be understood that
$\rho_0$ means $z_a q_a$ and that $P$ means $q_b$.  To see that
one can safely remove or insert such a sum between pairs of hard
emissions ($a$, $b$) without changing the underlying structure (the
`backbone') of the chain one observes that all
$\rhoti{\ell}<\qti{b}$. If $b$ is a $k$-conserving emission, then
so are all the soft emissions inserted before it. If $b$ is a
$k$-changing emission, then the soft emissions can also change $k$.
Two things prevent this from becoming a problem. Firstly, the
condition $\rhoti{\ell}<\qti{b}$ ensures that the rest of the chain
is not affected by the insertion. Secondly, the soft emissions are not
themselves affected by the value of $k$ (the non-Sudakov form factor
is always $1$, since $z=1$), so a change in $k$ has no effect on them.

Therefore in the calculation of the cross section one can simply use
\eqref{eq:CCFMdl}. The final state is then determined by inserting
sets of soft emissions \eqref{eq:softone} between every pair of hard
emissions (and also before the first one, and after the last
one).

\subsection{Pattern of soft emissions}

\FIGURE{    \input{soft-illus.pstex_t}\\
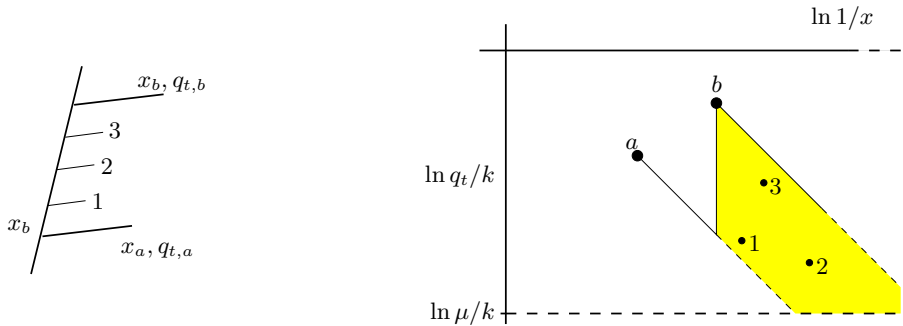
    \caption{Illustration of the insertion of soft emissions
      ($\ell=1,2,3$) between two hard emissions ($a,b$). The shaded
      area is the phase-space available to the soft emissions.}
    \label{fig:softillus}}

To understand the effect of soft emissions on the final state, let us
consider just the contribution from the soft emissions between two
hard emissions $a$ and $b$, as in figure~\ref{fig:softillus}. From
\eqref{eq:softone} the soft emissions are uniformally distributed,
with mean density
\begin{equation}
\label{eq:softdens}
\left\langle \frac{dn}{d\ln \qt \, d \ln 1/x}\right\rangle  = 2\asb
\end{equation}
in the shaded region, whose shape is determined as follows: a soft
emission $\ell$ has a momentum fraction $x_\ell = (1-\zeta_\ell) x_b
\ll x_b$, leading to the vertical boundary; the horizontal lower
boundary comes form the collinear cutoff: $\rhoti{\ell} > \mu$;
finally the two diagonal boundaries come from the angular ordering
conditions:
\begin{equation}
\qti{a}\frac{x_b}{x_a} <  \rho_i = (1-z_i) \rhoti{i} < \qti{b}
\end{equation}
where, as before, factors of $1-z_a$ and $1-z_b$ have been
approximated as $1$, with the error being subleading.

One notes that in the region where the soft emissions are allowed, the 
density of emissions \eqref{eq:softdens} is the same as the BFKL
density \eqref{eq:bfkldens}.

\subsection{Combination of soft and hard emissions}

\label{sec:sh}

\FIGURE{    \input{soft+hard_dk.pstex_t}\\
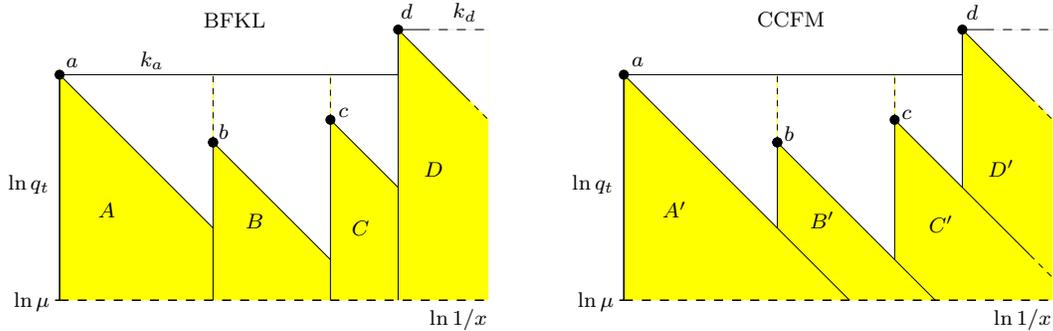
    \caption{DL equivalence between BFKL and CCFM final states. In the
      BFKL case, the black discs are those emissions which form an
      angular-ordered set. The shaded regions contain the remaining,
      unordered, emissions, which are independent, with mean density
      $2\asb$. In the CCFM case, the black discs are hard emissions,
      while the shaded regions contain the soft emissions, which are
      independent, with mean density $2\asb$.}
    \label{fig:softhard}}

Here we will see that the BFKL pattern of emissions is identical to
that from the CCFM (hard $+$ soft) equation. The fundamental point in
determining this equivalence will be that the ``order of the
emissions'' is not an observable property. It is only their final
distribution in rapidity and transverse momentum that matters.

Let us consider the first part of a chain, containing two $k$-changing
emissions, $a$ and $d$ (figure~\ref{fig:softhard}). These are
identical in BFKL and CCFM, from the results of
section~\ref{sec:divDL}. Emission $a$ has $x=1$.

In the BFKL case, the distribution for the next $k$-conserving
emission that is ordered in angle with respect to the first
  one (call it $b$) is given by
\begin{equation}
  \label{eq:NextAngBfkl}
  2\asb \frac{d\qti{b}}{\qti{b}} 
  \Delta(x_b,k_a,\qti{a})
  \Theta\left(\qti{b} - x_b \qti{a} \right).
\end{equation}
The non-Sudakov form factor arises simply through a calculation of the
probability of there not having been an angular ordered emission with
momentum fraction larger than $x_b$.  Equation \eqref{eq:NextAngBfkl}
is identical to the distribution for the next hard ($k$-conserving)
emission in the CCFM case.

Analogously, it is straightforward to determine that the distribution
of the next $k$-conserving angular-ordered emission (in the BFKL
case), $c$, is identical to that of the next hard emission (CCFM); and
similarly for the probability of there being no angular-ordered (hard)
$k$-conserving emissions in the BFKL (CCFM) case, before $d$.

So far we have only accounted for some of the emissions: in the BFKL
case there are the emissions between $a$ and $b$ which are not
angularly ordered with respect to $a$. They occupy the shaded region
labelled $A$; the emissions between $b$ and $c$ that are not angularly
ordered with respect to $b$ occupy the region $B$, and so on. These
emissions are independent and have the usual mean density
\eqref{eq:bfkldens}.

In the CCFM case, there are the soft emissions.  Those that come
before $a$ occupy the region labelled $A'$ (they are independent, with
mean density \eqref{eq:softdens}). The soft emissions between $a$ and
$b$ occupy region $B'$, those between $b$ and $c$, the region $C'$,
and so forth.

The sub-division of the ``non-ordered regions'', $A$--$D$ in the BFKL
case, and the soft regions $A'$--$D'$ in the CCFM case is different.
But the combination of $A$--$D$ is identical to the that of
$A'$--$D'$. Hence CCFM (hard $+$ soft) gives the same pattern of
emissions as BFKL.

\FIGURE{    \input{sh-end.pstex_t}\\
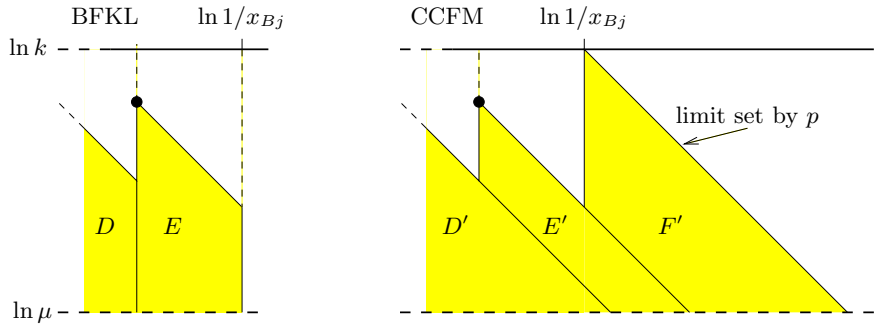
    \caption{A comparison between the end of a BFKL and the end of a
      CCFM chain. The CCFM chain has the maximum-angle limit set by
      $p=k$. As before, the shaded region contains independent
      emissions with mean density $2\asb$.}
    \label{fig:shend}}

This equivalence holds at the beginning and in the middle of the
chain. There is one place where a difference between BFKL and CCFM
double-logarithmic final states does arise, as illustrated in
figure~\ref{fig:shend}.

If BFKL evolution is carried out up to some limiting $x$, say
$x_{Bj}$, there are no emissions with $x_g < x_{Bj}$. This corresponds
to a vertical line cutting off the emissions at $x=x_{Bj}$ in
figure~\ref{fig:shend}. In the CCFM case, we will have the same
vertical line limiting the hard emissions. But the soft emissions are
delimited to the right only by diagonal lines (and by the intersection
with the collinear cutoff $\mu$).  In the particular case shown, where
the limiting angle is defined by $p=k$, the difference corresponds to
a triangle in $x,\qt$ space, containing on average $\asb \ln^2 p/\mu$
independent emissions. 

This difference is formally subleading since it does not contain any
$\ln x_{Bj}$ factors. In section~\ref{sec:BCfs} we saw that formally
subleading DL corrections can get promoted to affect the leading DLs.
This is not the case here though, since the effects of the difference
are confined to one end of the evolution chain, and so get
proportionately less important as one increases $\ln x_{Bj}$. For
example the total multiplicity in a chain where $k$ is determined by
the first (hard) emission is $2\asb \ln 1/x_{Bj} \ln k/\mu$ in the
BFKL case and $\asb (2\ln 1/x_{Bj} \ln k/\mu + \ln^2 k/\mu)$ in the
CCFM case ($p=k$). The difference is of relative order $\cO{\ln k /\mu
  / \ln 1/x_{Bj}}$ and hence negligible.

Depending on the kind of initial condition, it is also possible for
such differences to arise at the beginning of the chain. Given a
suitably perverse initial condition they can even be of the order of a
single $\as\ln^2x$ term (by cutting out an initial triangle of area
$\ln^2x$) --- however, again, they do not resum, and so do not give
rise to a whole series of double logarithms.

\section{Single logarithmic accuracy}
\label{sec:SingLog}

In this section we refine the techniques used above, in order to
demonstrate that BFKL and CCFM final states are equivalent at leading
(single) logarithmic accuracy.

The source of error in the above sections comes from the definition of
a $k$-conserving emission as any emission having $\qti{i} < k_{i-1}$.
In reality a $k$-conserving emission should have $\qti{i} \ll
k_{i-1}$. Thus there is a region of phase-space for each emission, of
size $\cO{\ln 1/x \ln 1/\eps}$, which is mistreated by the removal
emissions which are not quite $k$-conserving. Here, the parameter
$\eps$ has been introduced to define what is meant by $\qti{i} \ll
k_{i-1}$, namely $\qti{i} < \eps k_{i-1}$. Taking into account the
emission density proportional to $\asb$, one sees that for each
emission one is mistreating a contribution of $\cO{\asb \ln 1/x \ln
  1/\eps}$, which is LL.

\FIGURE{    \input{bb.pstex_t}\\
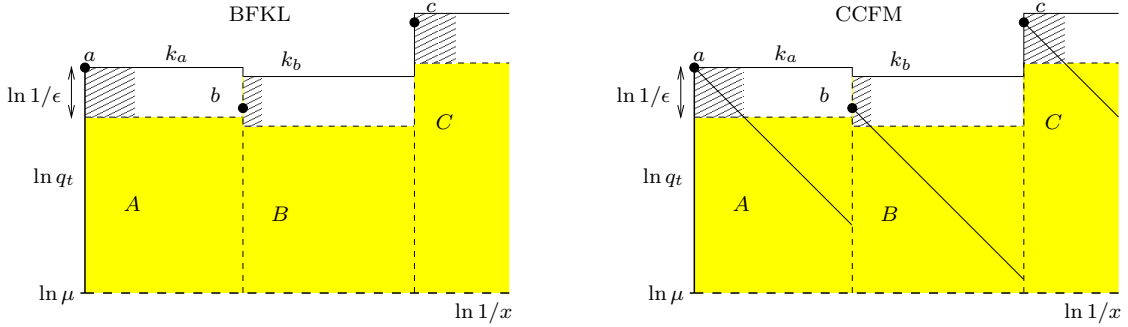
    \caption{Breaking BFKL and CCFM evolution into a backbone plus
      $k$-conserving emissions --- LL accuracy. Black dots are
      backbone emissions (the diagonal lines extending from them in
      the CCFM case indicate the angular ordering constraint for
      subsequent emissions). The shaded regions contain $k$-conserving
      emissions. The hashed regions indicate where the BFKL and CCFM
      emission densities may differ significantly (they should be
      understood to extend to $\qt \to \infty$).}
    \label{fig:bb}}

To extend the accuracy to be leading-logarithmic, one should therefore
repeat the analysis of the above sections using a proper definition of
a $k$-conserving emission. The basic procedure, as before, will be to
divide the emissions into two sets: a `backbone', consisting of those
emissions that are not $k$-conserving, and which therefore may affect
to the cross section.  And those that are $k$-conserving, and which
affect only the final state. The proof of the equivalence of BFKL and
CCFM final states then relies firstly on the BFKL and CCFM ensembles
of backbones being identical to LL accuracy; and secondly on the BFKL
or CCFM rules for the addition of the $k$-conserving emissions to a
given backbone having the same effect.

In the BFKL case, the backbone of $k$-changing emissions ($\qti{i}
> \eps k_{i-1}$) is given by
\begin{multline}
\label{eq:BFKLbb} 
  \cF(x,k) = \sum_{n=0}^\infty \prod_{i=1}^n \left(\asb
    \int\frac{dz_i}{z_i} \int_\mu\frac{d^2\vqti{i}}{\pi\qti{i}^2} 
  \Theta(\qti{i} - \eps k_{i-1}) \Delta_{\eps k_i} (z_i,k_i)\right)  \\
\delta\!\!\left(x - \prod_{i=1}^n z_i \right)
\delta^2\!\!\left(\vk + \sum_{i=1}^n \vqti{i} \right),
\end{multline}
where the cutoff has been introduced also in the form factor:
\begin{equation}
\ln  \Delta_{\eps k} (z,k) = -2\asb \ln \frac1z \ln
\frac{k}{\max(\mu,\eps k)}\,. 
\end{equation}
A sum over $k$-conserving emissions is then introduced after
each backbone emission: 
\begin{equation}
\label{eq:BFKLbbone}
  1 = 
  \sum_{m=0}^\infty \,\prod_{\ell=1}^{m}
  \left(\asb \int \frac{d\zeta_\ell}{\zeta_\ell} \int_\mu^{\eps k_i}
    \frac{d^2\vrhoti{\ell}}{\pi\rhoti{\ell}^2} 
    \Delta^{(\eps k_i )}(\zeta_\ell,k_i)
  \right) \Delta^{(\eps k_i)}(z_i/\zeta_\Pi, k_i)
  \,\Theta\!\!\left(\zeta_\Pi - z \right),
\end{equation}
with $\zeta_\Pi$ defined in analogy with \eqref{eq:zetaPi} and 
\begin{equation}
\Delta^{(\eps k)} (z,k) = \frac{\Delta(z,k)}{\Delta_{\eps k} (z,k)}\,.
\end{equation}
It is safe to neglect the contribution of the $\rho_\ell$'s to the
vector sum in \eqref{eq:BFKLbb} because their introduction can be
compensated for by modifying each of the $\qti{i}$ by a relative
amount $\cO{\eps}$. The $k$-conserving emissions fill up the shaded
regions in figure~\ref{fig:bb}, with the usual density of $2\asb$
independent emissions per unit of rapidity and $\ln \qt$.

In the CCFM case, let us first consider a backbone with just hard
emissions (even though some soft emissions might be more naturally
classified as belonging to the backbone): 
\begin{multline}
\label{eq:CCFMbb}
\cA(x,k,p) = \sum_{n=0}^\infty \prod_{i=1}^n \bigg(\asb
  \int\frac{dz_i}{z_i} \int \frac{d^2\vq_i}{\pi q_i^2}
  \,\Delta_{\eps k_i}(z_i,k_i,q_i)  \,\Theta(q_i - \eps
  k_{i-1})\,\\ \Theta(q_i - z_{i-1} 
  q_{i-1})\,\Theta(\qti{i} - \mu)\bigg) \cdot \Theta(p-z_n q_n)\,
\delta\!\left(x - \prod_{i=1}^n z_i \right) \delta^2\!\!\left(\vk +
  \sum_{i=1}^n \vqti{i} \right).
\end{multline}
The modified non-Sudakov form factor is
\begin{equation}
 \ln \Delta_{\eps k} (z,k,q) = 
 -2\asb \int_z^1 \frac{d\zeta}{\zeta} \int \frac{d\rho}{\rho}
  \,\Theta(k-\rho)  \, \Theta(\rho -\zeta q) \,\Theta(\rho - \mu) 
  \,\Theta(\rho - \eps k)\,.
\end{equation}
We then note that for $z < q/\eps k$ the BFKL and CCFM form factors
differ only by a constant subleading factor,
\begin{equation}
  \label{eq:BCff}
  \Delta_{\eps k}(z,k) = \Delta_{\eps k}(z,k,q) \cdot
  \exp\left(-\asb \ln^2 \frac{\qt}{\eps k} \right),
\end{equation}
and that for $z_{i} > \qti{i} /\eps k_{i}$ the phase-space limits for
the $\qti{i+1}$ integration become equal in the BFKL and CCFM cases,
namely 
\begin{equation}
  \frac{d^2\vqti{i+1}}{\pi\qti{i+1}^2} \,\Theta(\qti{i+1} - \eps k_i)\,.
\end{equation}
So we are now in a position to show that the BFKL and CCFM equations
lead to ensembles of backbones which are identical at LL accuracy:
from both ensembles we remove backbones containing branchings with
$z_{i} > \qti{i} /\eps k_{i}$ (i.e.\ %
emissions falling into the hashed regions of figure~\ref{fig:bb}).
Given that typically\footnote{In cases where $\qti{i} \gg k_i$ there
  is an additional subleading, but double-logarithmic price to pay,
  $\sim \asb \ln^2 \qti{i}/k_i$. In cases where $\qti{i+1} \gg k_i$
  the removed region will correspond to a correction $\sim \asb \ln
  \eps \ln \qti{i+1}/k_i$, which is also subleading.}%
$\qti{i} \sim k_i$, the likelihood of a branching violating this
condition contains a factor $\asb \ln^2 \eps$. Hence a subleading
fraction of backbones in each ensemble is removed. There is a
one-to-one correspondence between the remaining backbones in the BFKL
and CCFM ensemble. For a given backbone, the associated weights differ
between BFKL and CCFM due to the differences between the form factors,
\eqref{eq:BCff}, but again only by a subleading amount.

One might worry that since $\eps$ is a small parameter, $\asb \ln^2
\eps$ may not be a very `respectable' subleading correction. The other 
source of inaccuracy is corrections of $\cO{\eps}$. So it suffices to
take $\eps = \asb$ for both quantities to be truly subleading.

That the backbones should be the same at LL level could have been
guessed at right from the start, since the ensemble of backbones is
responsible for the determining the cross section, and we know that
the BFKL and CCFM cross sections differ only at subleading level.

The next stage in the study of the CCFM final state is the addition of
the $k$-conserving emissions.  The pattern for the hard $k$-conserving
emissions inserted after every backbone emissions is much as in
\eqref{eq:CCFMone},
\begin{multline}
\label{eq:CCFMbbone}
  1 = 
  \sum_{m=0}^\infty \,\prod_{\ell=1}^{m}
  \left(\asb \int \frac{d\zeta_\ell}{\zeta_\ell} \int^{\eps k_i}
    \frac{d^2\vrhoi{\ell}}{\pi\rho_{\ell}^2} 
    \,\Delta^{(\eps k_i)}(\zeta_\ell,k_i,\rho_\ell) \,\Theta(\rho_\ell -
    \zeta_{\ell-1} \rho_{\ell-1})  
    \,\Theta(\rho_{\ell} - \mu)
  \right) \\ \cdot \Delta^{(\eps k_i)}(z_i/\zeta_\Pi, k_i, \rho_0)
  \,\Theta\!\!\left(\zeta_\Pi - z_i \right),
\end{multline}
with the appropriate modification of the upper limit and of the
non-Sudakov form factor,
\begin{equation}
\Delta^{(\eps k)} (z,k,q) = \frac{\Delta(z,k,q)}{\Delta_{\eps k}
  (z,k,q)}\,. 
\end{equation}
The soft emissions are then to be inserted before each hard emission
$i$, and after the last one (in which case $q_i \equiv p$)
\begin{multline}
\label{eq:softbbone}
  1  - \cO{\asb \ln^2\eps} = \sum_{m=0}^\infty \prod_{\ell=1}^m
  \bigg(\asb 
    \int    \frac{d^2\vrho_\ell}{\pi \rho_\ell^2}  
    \frac{d\zeta_\ell}{1-\zeta_\ell}
  \frac{\Delta_S(\rho_\ell)}{\Delta_S( \rho_{\ell-1})} 
  \,\Theta(\rho_\ell - \rho_{\ell-1}) \\
  \,\Theta(\eps k_j - \rho_\ell)
  \,\Theta(\rhoti{\ell} -
  \mu)\bigg)  \cdot
  \frac{\Delta_S(p)}{\Delta_S(\rho_{n})} \Theta(q_{i}- \rho_m) ,
\end{multline}
where $\rho_0\equiv z_{i-1} q_{i-1}$. For a soft gluon with momentum
fraction $x_s$, the index $j$ (of $k_j$) is given by the condition
$x_{j-1} > x_s > x_j$.  This leads to a slightly different condition
from simple $k$-conservation: it ensures that the region in which soft
emissions can be present is at most the shaded region of
figure~\ref{fig:bb}.  This together with the limits on the $z$'s in
the backbone ensures that a soft emission does not affect the $k$
after the next hard emission by more than a relative amount $\eps$,
and that the soft emissions do not occupy the hashed regions in
figure~\ref{fig:bb}.  There is the price of a subleading contribution
$\cO{\asb \ln^2 \eps}$ (corresponding to the area of the lower
triangles in the hashed region in the CCFM diagram of
figure~\ref{fig:bb}) which arises from the incomplete cancellation
between the real and virtual parts of \eqref{eq:softbbone}, and which
should be included in the weight of the backbone. At LL accuracy this
is of no relevance.

Finally one needs to show that the combination of the $k$-conserving
hard and soft emissions fills up the shaded region in
figure~\ref{fig:bb} with a mean density of $2\asb$ independent
emissions per unit rapidity and $\ln \qt$, as in the BFKL case. The
procedure for doing so is identical to that used in
section~\ref{sec:sh}, but with appropriate modifications of the limits
on the emitted transverse momenta. There is therefore no need to
reproduce the explicit proof here.

This completes the demonstration that BFKL and CCFM final states are
identical at leading (single) logarithmic level.
 

\section{Conclusions}
\label{sec:conclusions}

The recurrent theme in this article has been that to study final-state
properties it is useful to split emissions into those which change the
exchanged transverse momentum (`backbone' emissions), and those which
do not. The former, being responsible for determining the cross
section, are almost bound to have the same pattern, since the BFKL and
CCFM cross sections are identical at LL order.

It is in the treatment of the latter, the collinear emissions, that
the BFKL and CCFM approaches at first sight appear as if they will
lead to different results. In the CCFM case there are two types of
collinear emissions, `hard' ($z\to 0$) and `soft' ($z\to 1$) ones.
Only after their combination does one obtain collinear emissions with
the same pattern as in BFKL.  Differences seen in the literature
between BFKL and CCFM final states were due to the inclusion of only
the hard collinear emissions (i.e.\ soft CCFM emissions cancel the
double logarithms of $x$ from hard CCFM emissions).

It should be emphasised that it is only their leading-logarithmic
predictions and not the BFKL and CCFM equations themselves (in the
sense of their physical content) that are equivalent: the BFKL
equation is derived in the limit of strong ordering in $x$ (without
coherence or soft emissions) --- this is then somewhat arbitrarily
extended to exact ordering.  The CCFM derivation deals explicitly with
the issues of coherence and soft radiation, as is necessary in order
to guarantee the leading-logarithms of the final state.

A consequence of these differences is, for example, that in the BFKL
equation, it is impossible to consider $z$ in the usual DGLAP sense,
since its value is largely determined, through the form factor, by the
value of the collinear cutoff --- for small cutoffs, $z$ is close $1$.
The structure of the CCFM equation is much more amenable to a direct
physical interpretation ($z$ does have the usual DGLAP
interpretation), and consequently perhaps a better starting point for
the correct inclusion of subleading effects \cite{NLL,NLLres} such as
the full splitting function. As to whether or not this is the case
will depend on whether other next-to-leading contributions can be
correctly included and resummed (for discussions of important physical
issues that need to be dealt with, the reader is referred to
\cite{NLLimprvd}).

\acknowledgments 

I would like to thank Pino Marchesini and St\'ephane Munier for a
series of helpful discussions. I am grateful also to Marcello
Ciafaloni, Jeffrey Forshaw, Misha Ryskin, Agoston Sabio-Vera and Bryan
Webber for useful conversations and comments.



\begin{thebibliography}{99}
\bibitem{BFKL} L.N. Lipatov, Sov. J. Phys. 23 (1976) 338;\\
       E.A. Kuraev, L.N. Lipatov and V.S. Fadin, Sov. Phys. JETP 45
       (1977) 199; \\
       Ya. Balitskii and L.N. Lipatov, Sov. J. Nucl. Phys. 28 (1978)
       822.
\bibitem{CCFMa}
  M. Ciafaloni, \npb{296}{1987}{249}.
\bibitem{CCFMb}
  S. Catani, F. Fiorani and G. Marchesini, \plb{234}{1990}{339};\\
  S. Catani, F. Fiorani and G. Marchesini, \npb{336}{1990}{18}.
\bibitem{Pino95} G. Marchesini, \npb{445}{1995}{40} [\hepph{9412327}].
\bibitem{FS}  J.R. Forshaw and A. Sabio Vera, \plb{B440}{1998}{141}
  [\hepph{9806394}]. 
\bibitem{Webber} B.R. Webber,  \plb{444}{1998}{81} [\hepph{9810286}].
\bibitem{FSW} J.R. Forshaw, A. Sabio Vera and B.R. Webber,
 Presented at 3rd UK Phenomenology Workshop on HERA Physics, Durham,
 England, September 1998, \hepph{9812318}.
\bibitem{EW} C. Ewerz and B.R. Webber, Cavendish-HEP-99/02.
\bibitem{DGLAP} V.N. Gribov and L.N. Lipatov, \sjnp{15}{1972}{438};\\
  G. Altarelli and G. Parisi, \npb{126}{1977}{298}; \\ Yu.L. Dokshitzer,
  \jetp{46}{1977}{641}. 
\bibitem{BMSS97} G. Bottazzi, G. Marchesini, G.P. Salam and
  M. Scorletti, \npb{505}{1997}{366} [\hepph{9702418}].
\bibitem{SMALLX}  G. Marchesini and B.R. Webber, \npb{386}{1992}{215}.
\bibitem{GBetal}  K. Golec-Biernat, L. Goerlich and J. Turnau,
 \npb{527}{1998}{289} [\hepph{9712345}].
\bibitem{NLL} 
  L.N. Lipatov and V.S. Fadin, \sjnp{50}{1989}{712}; \\
  V.S. Fadin, R. Fiore and M.I. Kotsky, \plb{539}{1995}{181}; \\ 
  V.S. Fadin, R. Fiore and M.I. Kotsky, \plb{387}{1996}{593}
  [\hepph{9605357}]; \\ 
  V.S. Fadin, and L.N. Lipatov, \npb{406}{1993}{259}; \\ V.S. Fadin,
  R. Fiore and A. Quartarolo, \prd{50}{1994}{5893} [\hepth{9405127}];
  \\ V.S. Fadin, 
  R. Fiore, and M.I. Kotsky, \plb{389}{1996}{737} [\hepph{9608229}];\\
  V.S. Fadin and L.N. Lipatov, \npb{477}{1996}{767} [\hepph{9602287}]; \\
  V.S. Fadin, M. I. Kotsky and L.N. Lipatov, \plb{415}{1997}{97}; \\
  S. Catani, M. Ciafaloni and F.Hautman, \plb{242}{1990}{97}; \\
  S. Catani, M. Ciafaloni and F.Hautman, \npb{366}{1991}{135}; \\
  V.S. Fadin, R. Fiore, A. Flashi, and M.I. Kotsky,
  \plb{422}{1998}{287} [\hepph{9711427}];\\ 
  V. Del Duca, \prd{54}{1996}{989};\\
  V. Del Duca, \prd{54}{1996}{4474};\\
  V. Del Duca and C.R. Schmidt, \prd{57}{1998}{4069} [\hepph{9711309}];\\
  V. Del Duca and C.R. Schmidt, \hepph{9810215}.

\bibitem{NLLres} M. Ciafaloni and G. Camici, \plb{430}{1998}{127}
  [\hepph{9803389}];\\ 
  M. Ciafaloni, \plb{429}{1998}{349} [\hepph{9801322}];\\
  M. Ciafaloni and G. Camici, \plb{412}{1997}{396} [\hepph{9707390}];\\
  V.S. Fadin and L.N. Lipatov, \plb{429}{1998}{127} [\hepph{9802290}].

\bibitem{NLLimprvd} B. Andersson, G. Gustafson and J. Samuelsson,
  \npb{467}{1996}{443};\\
  D.A. Ross, \plb{431}{1998}{161} [\hepph{9804332}];\\
  Yuri V. Kovchegov and A.H. Mueller, \plb{439}{1998}{428}
  [\hepph{}9805208];\\
  E. Levin, \hepph{9806228};\\
  N. Armesto, J. Bartels and M.A. Braun, \plb{442}{1998}{459}
  [\hepph{9808340}];\\ 
  G.P. Salam, \jhep{9807}{1998}{019} [\hepph{9806482}];\\
  M. Ciafaloni and D. Colferai, \hepph{9812366};\\
  Stanley J. Brodsky, Victor S. Fadin, Victor T. Kim, Lev N. Lipatov
  and Grigorii B. Pivovarov, \hepph{9901229};\\
  R.S. Thorne, \hepph{9901331}. 


\end{thebibliography}
\end{document}